\documentclass[preprint,prd]{revtex4}
\usepackage{amssymb}
\usepackage{amsmath}
\usepackage{graphicx}

\input{tcilatex}
\begin{document}

\title[]{Self-Inductance and the Mass of Current Carriers in a Circuit}
\author{Timothy H. Boyer}
\affiliation{Department of Physics, City College of the City University of New York, New
York, New York 10031}

\begin{abstract}
In this article, the self-inductance of a\ circular circuit is treated from
an untraditional, particle-based point of view. \ The electromagnetic fields
of Faraday induction are calculated explicitly from the point-charge fields
derived from the Darwin Lagrangian for particles confined to move in a
circular orbit. \ For a one-particle circuit (or for $N$ \textit{%
non-interacting} particles), the induced electromagnetic fields depend upon
the mass and charge of the current carriers while energy is transferred to
the kinetic energy of the particle (or particles). \ However, for an
interacting multiparticle circuit, the mutual electromagnetic interactions
between particles can dominate the behavior so that the mass and charge of
the individual particles becomes irrelevant; the induced fields are then
comparable to the inducing fields and energy goes into magnetic energy. \ In
addition to providing a deeper understanding of self-inductance, the example
suggests that the claims involving \textquotedblleft hidden mechanical
momentum\textquotedblright\ in connection with momentum balance for
interacting multiparticle systems are unlikely to be accurate.
\end{abstract}

\maketitle

\subsection{Introduction}

When the self-inductance of a circuit is discussed in electromagnetism
textbooks,\cite{Griffiths}\cite{Jackson} the masses and the charges of the
current carriers in the circuit are never mentioned. \ Yet clearly, the
charge carriers play a crucial role in the circuit's electromagnetic
behavior. \ Why is it that the self-inductance of a circuit can be discussed
without any mention of the details of the charge carriers? \ 

In this article, we take an untraditional viewpoint and treat
self-inductance as arising from the electromagnetic fields of point charges
as derived from the Darwin Lagrangian.\cite{unaware} \ In the analysis, we
illustrate the roles of particle mass and particle charge in a simple
electromagnetic situation. \ The analysis makes clear just when and why the
details of the charge carriers disappear from self-inductive behavior. \ The
unusual perspective presented here seems of interest not only as an analysis
providing a deeper understanding of electromagnetic theory, but also as a
suggestive commentary on the controversial topic of "hidden mechanical
momentum." \ Writings in the current literature of electromagnetism\cite%
{letter} (including some excellent text books,\cite{Griffiths}\cite{Jackson}%
) insist that the \textit{mechanical} momentum of the carriers of charge is
the basis for a "hidden mechanical momentum" which must be involved in the
momentum balance of electromagnetic systems. \ \ In contradiction to this
widely-quoted view, we will suggest that our example illustrates that
"hidden mechanical momentum" is a concept which is unlikely to play any role
in multiparticle electromagnetism. \ 

\subsection{Textbook Discussion of Self-Inductance}

According to the standard textbooks of electromagnetism, if an external emf $%
emf_{ext}$\ is present in a continuous circuit with a self-inductance $L$
and resistance $\mathcal{R}$, the current $i$ in the circuit is given by the
differential equation%
\begin{equation}
emf_{ext}=L\frac{di}{dt}+i\mathcal{R}  \label{e1}
\end{equation}%
Here the self-inductance $L$ is a quantity which depends upon the geometry
of the circuit. \ For example, the self-inductance per unit length $L/l$ of
a long solenoid\cite{solenoid} of $n$ turns per unit length and
cross-section area $\mathcal{A}$ is given in gaussian units by 
\begin{equation}
L/l=4\pi n^{2}\mathcal{A}/c^{2}  \label{e1a}
\end{equation}%
where $c$ is the speed of light in vacuum. \ The energy balance for the
circuit is found by multiplying Eq. (\ref{e1}) by the current $i$%
\begin{equation}
emf_{ext}\times i=\frac{d}{dt}\left( \frac{1}{2}Li^{2}\right) +i^{2}\mathcal{%
R}  \label{e2}
\end{equation}%
corresponding to a power $emf_{ext}\times i$ delivered by the external emf
going into the time-rate-of-change of magnetic energy $(1/2)Li^{2}$ stored
in the inductor and the power $i^{2}\mathcal{R}$ lost in the resistor. \ In
the standard textbook discussion, the mass of the current carriers is never
mentioned; it appears neither in the current calculation nor in the energy
balance. \ 

\subsection{Simple Model for a Particle-Based Analysis}

There is an alternative point of view regarding self-inductance which
directs attention to the charges which carry the current rather than to the
geometry of the electromagnetic circuit. \ The appearance of the speed of
light $c$ in the expression (\ref{e1a}) for the self-inductance per unit
length of a long solenoid reminds us that self-inductance involves a
relativistic effect, and hence our model must be relativistic at least
through order $v^{2}/c^{2}$. \ Accordingly, the analysis makes use of the
Darwin Lagrangian and the derived expressions for the electromagnetic fields
of interacting point charges through order $1/c^{2}.$ \ We will illustrate
this alternative point of view for the case of $N$ equally-spaced particles,
all of mass $m$ and charge $e,$ which are held by external centripetal
forces in a circular orbit of radius $R$ centered on the origin in the $xy$%
-plane. There is no frictional force and hence no resistance $\mathcal{R}$
in the model. \ Rather, the system may be thought of as consisting of
charged beads sliding on a frictionless ring. \ 

If an axially symmetric magnetic field is applied perpendicular to the plane
of the circular orbit in the $-\widehat{z}$-direction and is increasing in
magnitude, then the time-rate-of-change of the external magnetic field will
produce an external electric field $\mathbf{E}_{ext}$ in a circular pattern
in the $\widehat{\phi }$-direction $\mathbf{E}_{ext}\mathbf{(r})=\widehat{%
\phi }E_{ext}(r).$ The external emf around the circular orbit is given by 
\begin{equation}
emf_{ext}=\toint \mathbf{E}_{ext}\mathbf{(r})\cdot \mathbf{dr}=2\pi
RE_{ext}(R)  \label{e3}
\end{equation}%
The external electric field $\mathbf{E}_{ext}(\mathbf{r})$ places a
tangential force $\mathbf{F}_{i}=e_{i}\widehat{\phi }_{i}E_{ext}(R)$ on the $%
i$th particle located at $\mathbf{r}_{i}.$ \ The self-inductance of the
charged-particle system is determined by the response of all the particles $%
e_{i}$ in the circular orbit.

\subsection{Electromagnetic Fields of Particles from the Darwin Lagrangian}

In order to determine the behavior of the charged particles $e_{i}$, we turn
to the Darwin Lagrangian\cite{Darwin} which represents the behavior of
interacting charged particles through order $1/c^{2}$%
\begin{eqnarray}
\mathcal{L} &\mathcal{=}&\tsum\limits_{i=1}^{i=N}m_{i}c^{2}\left( -1+\frac{%
\mathbf{v}_{i}^{2}}{2c^{2}}+\frac{(\mathbf{v}_{i}^{2})^{2}}{8c^{4}}\right) -%
\frac{1}{2}\tsum\limits_{i=1}^{i=N}\tsum\limits_{j\neq i}\frac{e_{i}e_{j}}{|%
\mathbf{r}_{i}-\mathbf{r}_{j}|}  \notag \\
&&+\frac{1}{2}\tsum\limits_{i=1}^{i=N}\tsum\limits_{j\neq i}\frac{e_{i}e_{j}%
}{2c^{2}}\left[ \frac{\mathbf{v}_{i}\cdot \mathbf{v}_{j}}{|\mathbf{r}_{i}-%
\mathbf{r}_{j}|}+\frac{\mathbf{v}_{i}\cdot (\mathbf{r}_{i}-\mathbf{r}_{j})%
\mathbf{v}_{j}\cdot (\mathbf{r}_{i}-\mathbf{r}_{j})}{|\mathbf{r}_{i}-\mathbf{%
r}_{j}|^{3}}\right]  \notag \\
&&-\tsum\limits_{i=1}^{i=N}e_{i}\Phi _{ext}(\mathbf{r}_{i},t)+\tsum%
\limits_{i=1}^{i=N}e_{i}\frac{\mathbf{v}_{i}}{c}\cdot \mathbf{A}_{ext}(%
\mathbf{r}_{i},t)  \label{e4}
\end{eqnarray}%
where the last line includes the scalar potential $\Phi _{ext}$ and vector
potential $\mathbf{A}_{ext}$ associated with the external electromagnetic
fields. \ The Lagrangian equations of motion can be rewritten in the form of
Newton's second law $d\mathbf{p}/dt=d(m\gamma \mathbf{v})/dt=\mathbf{F}$
with $\gamma =(1-v^{2}/c^{2})^{-1/2}.$ \ In this Newtonian form, we have%
\begin{eqnarray}
&&\frac{d}{dt}\left[ \frac{m_{i}\mathbf{v}_{i}}{(1-\mathbf{v}%
_{i}^{2}/c^{2})^{1/2}}\right]  \notag \\
&\approx &\frac{d}{dt}\left[ m_{i}\left( 1+\frac{\mathbf{v}_{i}^{2}}{2c^{2}}%
\right) \mathbf{v}_{i}\right] =e_{i}\mathbf{E}+e_{i}\frac{\mathbf{v}_{i}}{c}%
\times \mathbf{B}  \notag \\
&=&e_{i}\left( \mathbf{E}_{ext}(\mathbf{r}_{i},t)+\tsum\limits_{j\neq i}%
\mathbf{E}_{j}(\mathbf{r}_{i},t)\right) +e_{i}\frac{\mathbf{v}_{i}}{c}\times
\left( \mathbf{B}_{ext}(\mathbf{r}_{i},t)+\tsum\limits_{j\neq i}\mathbf{B}%
_{j}(\mathbf{r}_{i},t)\right)  \label{e5}
\end{eqnarray}%
with the Lorentz force on the $i$th particle arising from the external
electromagnetic fields and from the electromagnetic fields of the other
particles. \ The electromagnetic fields due to the $j$th particle are given
through order $v^{2}/c^{2}$ by\cite{PageAdams}%
\begin{eqnarray}
\mathbf{E}_{j}(\mathbf{r,t}) &=&e_{j}\frac{(\mathbf{r}-\mathbf{r}_{j})}{|%
\mathbf{r}-\mathbf{r}_{j}|^{3}}\left[ 1+\frac{\mathbf{v}_{j}^{2}}{2c^{2}}-%
\frac{3}{2}\left( \frac{\mathbf{v}_{j}\cdot (\mathbf{r}-\mathbf{r}_{j})}{c|%
\mathbf{r}-\mathbf{r}_{j}|}\right) ^{2}\right]  \notag \\
&&-\frac{e_{j}}{2c^{2}}\left( \frac{\mathbf{a}_{j}}{|\mathbf{r}-\mathbf{r}%
_{j}|}+\frac{\mathbf{a}_{j}\cdot (\mathbf{r}-\mathbf{r}_{j})(\mathbf{r}-%
\mathbf{r}_{j})}{|\mathbf{r}-\mathbf{r}_{j}|^{3}}\right)  \label{e6}
\end{eqnarray}%
and 
\begin{equation}
\mathbf{B}_{j}(\mathbf{r},t)=e_{j}\frac{\mathbf{v}_{j}}{c}\times \frac{(%
\mathbf{r}-\mathbf{r}_{j})}{|\mathbf{r}-\mathbf{r}_{j}|^{3}}  \label{e7}
\end{equation}%
where in Eq. (\ref{e6}) the quantity $\mathbf{a}_{j}$ refers to the
acceleration of the $j$th particle.

\subsection{One-Particle Model for a Circuit}

\subsubsection{Motion of the Charged Particle}

We start with the case when there is only one charged particle of mass $m$
and charge $e$ in the circular orbit. \ In this case, the tangential
acceleration $a_{\phi }$ of the charged particle $e$ in the circular orbit
arises from the force of only the external electric field $\mathbf{E}_{ext}.$
\ From Eq. (\ref{e5}), written for a single particle and with $d(m\gamma
v)/dt=m\gamma ^{3}a_{\phi }$ where $\gamma =(1-v^{2}/c^{2})^{-1/2}$, we have 
\begin{equation}
a_{\phi }=\frac{eE_{ext}(R)}{m\gamma ^{3}}  \label{e8}
\end{equation}%
where $E_{ext}(R)$ is the magnitude of the tangential electric field due to
the external emf $emf_{ext}$ at the position of the charge $e.$

\subsubsection{Magnetic Field of the Charged Particle \ }

The magnetic field $\mathbf{B}_{e}$ at the center of the circular orbit due
to the accelerating charge $e$ is given by Eq. (\ref{e7}) 
\begin{equation}
\mathbf{B}_{e}(0,t)=\widehat{k}e\frac{v}{cR^{2}}  \label{e9}
\end{equation}%
where the velocity $v$ is increasing since the external electric field $%
\mathbf{E}_{ext}$ gives a positive charge $e$ a positive acceleration in the 
$\widehat{\phi }$-direction. This magnetic field $\mathbf{B}_{e}$ produced
by the orbiting charge $e$ is increasing in the $\widehat{z}$-direction,
which is in the opposite direction from the increasing external magnetic
field which created the external electric field $\mathbf{E}_{ext}(\mathbf{r}%
) $ and the external emf $emf_{ext}$ in Eq. (\ref{e3}). \ 

\subsubsection{Induced Electric Field}

Associated with this changing magnetic field $\mathbf{B}_{e}$ created by the
orbiting charge $e,$ there should be an induced electric field $\mathbf{E}%
_{e}(\mathbf{r},t)$ according to Faraday's law. \ Thus averaging over the
circular motion of the charge $e$, we expect an average induced tangential
electric field $\left\langle E_{e\phi }(r)\right\rangle $ at a distance $r$
from the center of the circular orbit (where $r<<R$ so that the magnetic
field $\mathbf{B}_{e}$ has approximately the value $\mathbf{B}(0,t)$ at the
center) given from Eq. (\ref{e9}) by 
\begin{eqnarray}
2\pi r\,\left\langle E_{e\phi }(r)\right\rangle &=&emf_{e}=-\frac{1}{c}\frac{%
d\Phi _{e}}{dt}=-\frac{1}{c}\frac{d}{dt}[B_{e}(0,t)\pi r^{2}]  \notag \\
&=&=-\frac{1}{c}\frac{d}{dt}\left[ e\frac{v}{cR^{2}}\pi r^{2}\right] =-\frac{%
1}{c}\left( e\frac{a_{\phi }}{cR^{2}}\right) \pi r^{2}  \label{eee10}
\end{eqnarray}%
since $dv/dt=a_{\phi .}$ \ Using Eq. (\ref{e8}), the average tangential
electric field follows from Eq. (\ref{eee10}) as%
\begin{equation}
\left\langle \mathbf{E}_{e\phi }(r,t)\right\rangle =-\widehat{\phi }\frac{%
e^{2}rE_{ext}(R)}{2mc^{2}\gamma ^{3}R^{2}}  \label{e11}
\end{equation}

We will now show that this induced average tangential electric field $%
\left\langle \mathbf{E}_{e\phi }(r,t)\right\rangle $ is exactly the average
electric field due to the charge $e$ obtained by use of the electric field
expression given in Eq. (\ref{e6}). \ Thus we assume that the charge $e$ is
located momentarily at $\mathbf{r}_{e}=\widehat{x}R\cos \phi _{e}+\widehat{y}%
R\sin \phi _{e}$ and average over the phase $\phi _{e}$. \ Since the entire
situation is axially symmetric when averaged over $\phi _{e},$ we may take
the field point along the $x$-axis at $\mathbf{r}=\widehat{x}r$, and later
generalize to cylindrical coordinates. \ The velocity fields given in the
first line of Eq. (\ref{e6}) point from the charge $e$ to the field point. \
Also, the velocity fields are even if the sign of the velocity $\mathbf{v}%
_{e}$ is changed to $-\mathbf{v}_{e}.$ \ Thus the velocity fields when
averaged over the circular orbit can point only in the radial direction. \
The accelerations fields arising from the centripetal acceleration of the
charge will also point in the radial direction. \ Since we are interested in
the average tangential component of the field $\mathbf{E}_{e}$, we need to
average over only the tangential acceleration terms in the second line of
Eq. (\ref{e6}). \ If the field point is close to the center of the circular
orbit so that $r<<R,$ then we may expand in powers of $r/R;$ we retain only
the first-order terms, giving $|\widehat{x}r-\mathbf{r}_{e}|^{-1}\approx
R^{-1}(1+\widehat{x}r\cdot \mathbf{r}_{e}/R^{2})$ and$|\widehat{x}r-\mathbf{r%
}_{e}|^{-3}\approx R^{-3}(1+3\widehat{x}r\cdot \mathbf{r}_{e}/R^{2}).$ \
Then the average tangential component of the electric field due to the
charge $e$ can be written as 
\begin{eqnarray}
\left\langle \mathbf{E}_{e\phi }(\widehat{x}r,t)\right\rangle
&=&\left\langle -\frac{e}{2c^{2}}\left( \frac{\mathbf{a}_{e\phi }}{|\widehat{%
x}r-\mathbf{r}_{e}|}+\frac{\mathbf{a}_{e\phi }\cdot (\widehat{x}r-\mathbf{r}%
_{e})(\widehat{x}r-\mathbf{r}_{e})}{|\widehat{x}r-\mathbf{r}_{e}|^{3}}%
\right) \right\rangle  \notag \\
&=&\left\langle -\frac{e}{2c^{2}}\left[ \frac{\mathbf{a}_{e\phi }}{R}\left(
1+\frac{\widehat{x}r\cdot \mathbf{r}_{e}}{R^{2}}\right) +\frac{\mathbf{a}%
_{e\phi }\cdot (\widehat{x}r-\mathbf{r}_{e})(\widehat{x}r-\mathbf{r}_{e})}{%
R^{3}}\left( 1+\frac{3\widehat{x}r\cdot \mathbf{r}_{e}}{R^{2}}\right) \right]
\right\rangle  \label{e12}
\end{eqnarray}%
Now we average over the phase $\phi _{e}$ with $\mathbf{r}_{e}=\widehat{x}%
R\cos \phi _{e}+\widehat{y}R\sin \phi _{e}$ and $\mathbf{a}_{e\phi
}=a_{e\phi }(\mathbf{-}\widehat{x}\sin \phi _{e}+\widehat{y}\cos \phi _{e}.$
\ We note that $\left\langle \mathbf{a}_{e\phi }\right\rangle =0,$ $\mathbf{a%
}_{e\phi }\cdot \mathbf{r}_{e}=0,$ $\left\langle \mathbf{a}_{e\phi }(%
\widehat{x}\cdot \mathbf{r}_{e})\right\rangle =\widehat{y}a_{e\phi
}R/2=-\left\langle (\mathbf{a}_{e\phi }\cdot \widehat{x})\mathbf{r}%
_{e}\right\rangle .$ \ After averaging and retaining terms through order $%
r/R $, equation (\ref{e12}) becomes%
\begin{equation}
\left\langle \mathbf{E}_{e\phi }(\widehat{i}r,t)\right\rangle =-\widehat{y}%
\frac{ea_{\phi }r}{2c^{2}R^{2}}  \label{e13}
\end{equation}%
which is in agreement with our earlier results in Eqs. (\ref{eee10}) and (%
\ref{e11}). \ 

\subsubsection{Limit on the Induced Electric Field}

We are now in a position to comment on the average response of our
one-particle circuit to the applied external emf. For this one-particle
example, the response depends crucially upon the mass $m$ and charge $e$ of
the particle. \ When the mass $m$ is large, the acceleration of the charge
is small; therefore the induced tangential electric field $\mathbf{E}_{e\phi
}$ in Eq. (\ref{e11}) is small. \ This large-mass situation is what is
usually assumed in examples of charged rings responding to external emfs.%
\cite{Feynman} \ On the other hand, if we try to increase the induced
electromagnetic field $\mathbf{E}_{e\phi }$ by making the mass $m$ small, we
encounter a fundamental limit of electromagnetic theory. \ The allowed mass $%
m$ is limited below by considerations involving the classical radius of the
electron $r_{cl}=e^{2}/(mc^{2}).$ \ Classical electromagnetic theory is
valid only for distances large compared to the classical radius of the
electron. \ Thus in our example where the radius $R$ of the orbit is a
crucial parameter, we must have $R>>r_{cl}.$ Thus we require the mass $%
m>>e^{2}/(Rc^{2})$ and so $e^{2}/(mc^{2}R)<<1.$ \ Combing this limit with $%
r/R<1,$ and $1<\gamma $ leads to a limit on the magnitude of the induced
electric field in Eq. (\ref{e11}) 
\begin{equation}
\left\langle E_{e\phi }(r,t)\right\rangle <<E_{ext}(R)\text{ \ \ for \ \ }r<R
\label{e14}
\end{equation}%
The induced electric field is small compared to the external electric field
associated with the external emf. \ 

\subsubsection{Energy Balance}

We also note that the power delivered by the external electric field goes
into kinetic energy of the orbiting particle. \ Thus if take the
Newton's-second-law equation giving Eq. (\ref{e8}) and multiply by the speed 
$v$ of the particle, we have%
\begin{equation}
\frac{d}{dt}(m\gamma v)v=\frac{d}{dt}(m\gamma c^{2})=m\gamma ^{3}a_{\phi
}v=eE_{ext}(R)v  \label{e15}
\end{equation}%
so that the power $eE_{ext}(R)v$ delivered to the charge $e$ by the external
electric field goes into kinetic energy of the particle. \ 

The situation of a one-particle circuit can be summarized as follows. \ For
the one-particle circuit, the induced electric field is small compared to
the external electric field and depends explicitly upon the particle's mass
and charge, while the energy transferred by the external field goes into
kinetic energy of the one charged particle. \ Clearly this is not the
situation which we usually associate with electromagnetic induction for
circuit problems.

\subsection{Multi-Particle Model for a Circuit}

\subsubsection{Motion of the Charged Particles}

In order to make contact with the usual discussion of self-inductance in
electromagnetism, we must go to the situation of many electric charges. \
The force on any charge in the circular orbit is now the sum of the forces
due to the original external electric field plus that due to the fields of
all the other charged particles in the circular orbit as given in Eq. (\ref%
{e5}). \ The magnetic force $e_{i}\mathbf{v}_{i}\times \mathbf{B}/c$ is
simply a deflection and does not contribute the the tangential acceleration.
\ Thus the equation of motion for the $i$th particle becomes%
\begin{eqnarray}
&&\frac{d}{dt}(m_{i}\gamma _{i}\mathbf{v}_{i})\cdot \widehat{\phi }  \notag
\\
&=&m_{i}\gamma _{i}^{3}a_{i\phi }=m_{i}\gamma _{i}^{3}R\frac{d^{2}\phi _{i}}{%
dt^{2}}  \notag \\
&=&\widehat{\phi }_{i}\cdot e_{i}\left\{ \mathbf{E}_{ext}\mathbf{(r}%
_{i})+\sum\limits_{j\neq i}e_{j}\frac{(\mathbf{r}_{i}-\mathbf{r}_{j})}{|%
\mathbf{r}_{i}-\mathbf{r}_{j}|^{3}}-\sum\limits_{j\neq i}\frac{e_{j}}{2c^{2}}%
\left( \frac{\mathbf{a}_{j}}{|\mathbf{r}_{i}-\mathbf{r}_{j}|}+\frac{\mathbf{a%
}_{j}\cdot (\mathbf{r}_{i}-\mathbf{r}_{j})(\mathbf{r}_{i}-\mathbf{r}_{j})}{|%
\mathbf{r}_{i}-\mathbf{r}_{j}|^{3}}\right) \right\}  \label{ee16}
\end{eqnarray}%
\qquad Since the particles are equally spaced is around the circular orbit
and all have the same charge $e$ and mass $m$, the situation is axially
symmetric. \ The equation of motion for every charge takes the same form,
and the angular acceleration of each charge is the same, $d^{2}\phi
_{i}/dt^{2}=d^{2}\phi /dt^{2}.$ \ For simplicity of calculation, we will
take the $N$th particle along the $x$-axis so that $\phi _{N}=0,$ $\mathbf{r}%
_{N}=\widehat{x}R,$ and $\widehat{\phi }_{N}=\widehat{y}.$ \ The other
particles are located at $\mathbf{r}_{j}=\widehat{x}R\cos (2\pi j/N)+%
\widehat{y}R\sin (2\pi j/N),$ corresponding to an angle $\phi _{j}=2\pi j/N$
for $j=1,2,...,N.\ $The tangential acceleration of the $j$th particle is
given by $\mathbf{a}_{j\phi }=(d^{2}\phi /dt^{2})[-\widehat{x}R\sin (2\pi
j/N)+\widehat{y}R\cos (2\pi j/N)].$ \ By symmetry, it is clear that the
electrostatic fields, the velocity fields, and the centripetal acceleration
fields of the other particles can not contribute to the tangential electric
field at particle $N.$ \ The equation of motion for the tangential
acceleration for each charge in the circular orbit is the same as that for
the $N$th particle, which from Eq. (\ref{ee16}) is 
\begin{equation}
m\gamma ^{3}R\frac{d^{2}\phi }{dt^{2}}=\left\{
eE_{ext}(R)-\sum\limits_{j=1}^{j=N-1}\frac{e^{2}}{2c^{2}}\left( \frac{%
\widehat{y}\cdot \mathbf{a}_{j\phi }}{|\widehat{x}R-\mathbf{r}_{j}|}+\frac{%
\mathbf{a}_{j\phi }\cdot (\widehat{x}R-\mathbf{r}_{j})\widehat{y}\cdot (%
\widehat{x}R-\mathbf{r}_{j})}{|\widehat{x}R-\mathbf{r}_{j}|^{3}}\right)
\right\}  \label{e16a}
\end{equation}%
Now we evaluate the distance between the $j$th particle and the $N$th
particle in the circular orbit as 
\begin{equation}
|\widehat{x}R-\mathbf{r}_{j}|=[2R^{2}-2R^{2}\cos (2\pi
j/N)]^{1/2}=[4R^{2}\sin ^{2}(\pi j/N)]^{1/2}=|2R\sin (\pi j/N)|  \label{e17a}
\end{equation}%
while 
\begin{equation}
\widehat{y}\cdot \mathbf{a}_{j\phi }=(d^{2}\phi /dt^{2})R\cos (2\pi j/N)
\label{e17b}
\end{equation}%
and 
\begin{equation}
\mathbf{a}_{j}\cdot (\widehat{x}R-\mathbf{r}_{j})\widehat{y}\cdot (\widehat{x%
}R-\mathbf{r}_{j})=(d^{2}\phi /dt^{2})R[-R\sin (2\pi j/N)][-R\sin (2\pi j/N)]
\label{e17c}
\end{equation}%
\ Then equation equation (\ref{e16a}) becomes%
\begin{eqnarray}
&&m\gamma ^{3}R\frac{d^{2}\phi }{dt^{2}}  \notag \\
&=&eE_{ext}(R)-\frac{d^{2}\phi }{dt^{2}}\sum\limits_{j=1}^{j=N-1}\frac{e^{2}%
}{2c^{2}}\left( \frac{R\cos (2\pi j/N)}{|2R\sin (\pi j/N)|}+\frac{R[R\sin
(2\pi j/N)][R\sin (2\pi j/N)]}{|2R\sin (\pi j/N)|^{3}}\right)  \label{ee18}
\end{eqnarray}%
\qquad or, solving for $d^{2}\phi /dt^{2},$%
\begin{equation}
\frac{d^{2}\phi }{dt^{2}}=eE_{ext}(R)\left[ m\gamma
^{3}R+\sum\limits_{j=1}^{j=N-1}\frac{e^{2}}{2c^{2}}\left( \frac{\cos (2\pi
j/N)}{|2\sin (\pi j/N)|}+\frac{[\sin ^{2}(2\pi j/N)]}{|2\sin (\pi j/N)|^{3}}%
\right) \right] ^{-1}  \label{e19}
\end{equation}%
If there is only one particle on the frictionless ring so that $N=1$, the
sum disappears, and the tangential acceleration corresponds to the result
obtained earlier in Eq. (\ref{e8}) above with $Rd^{2}\phi /dt^{2}=a_{\phi }$%
. \ We note that the mass term in Eq. (\ref{e19}) remains unchanged by the
number of particles while the sum increases with each additional particle. \
Thus if there are many particles, then the electric field at particle $i$
due to the other particles $j$\ can lead to so large a sum in Eq. (\ref{e19}%
) that the mass contribution $m\gamma ^{3}R$ becomes insignificant. \ In
this case, the common angular acceleration of each particle becomes from Eq.
(\ref{e19})%
\begin{equation}
\frac{d^{2}\phi }{dt^{2}}\approx \frac{2c^{2}}{e}E_{ext}(R)\left[
\sum\limits_{j=1}^{j=N-1}\left( \frac{\cos (2\pi j/N)}{|2\sin (\pi j/N)|}+%
\frac{[\sin ^{2}(2\pi j/N)]}{|2\sin (\pi j/N)|^{3}}\right) \right] ^{-1}
\label{e20}
\end{equation}%
\ We see that in this multiparticle situation the angular acceleration no
longer depends upon the mass $m$ of the charge carriers.

\subsubsection{Induced Electric Field}

Furthermore, in this multiparticle situation where the particle mass becomes
insignificant, the left-hand side of Eq. (\ref{e16a}) is negligible, so that
the sum $\tsum_{j}\mathbf{E}_{ej}(\mathbf{r}_{i})~$of the acceleration
fields of all the other charges $e_{j}$ cancels the external electric field $%
\mathbf{E}_{ext}(\mathbf{r}_{i})$ of the external emf at the position $%
\mathbf{r}_{i}$\ of each charge in the circular orbit

\begin{equation}
-\mathbf{E}_{ext}\mathbf{(r}_{i}\mathbf{)\approx E}_{e}(\mathbf{r}%
_{i})=\tsum_{j\neq i}\mathbf{E}_{ej}(\mathbf{r}_{i})  \label{e21}
\end{equation}%
Thus analogous to the situation in electrostatics where the fields of the
charges in a conductor move to new positions so as to cancel the external
electric field at the position of each charge, here the charges accelerate
so that the acceleration fields cancel the external field at the position of
each charge. \ Now the induced electric field $\mathbf{E}_{e}(\mathbf{r}%
)=\tsum_{i}\mathbf{E}_{ei}(\mathbf{r})$ at a general field point due to the
orbit particles is independent of the charge $e$ of the charge carriers,
since the angular acceleration in Eq. (\ref{e20}) depends inversely as the
charge $e,$ and this inverse dependence upon $e$ cancels with the $e$
appearing in Eq. (\ref{e6}) so as to give an induced electric field which is
independent of the charge on the charge carriers. \ 

\subsubsection{Self-Inductance of the Circuit}

The self-inductance of the multiparticle circuit can be obtained from Eq. (%
\ref{e1}) when the circuit resistance vanishes. If the resistance vanishes,
the external emf $emf_{ext}=2\pi RE_{ext}(R)$ around the ring equals the
self-inductance multiplied by the time-rate-of-change of the current $%
i=Ne(d\phi /dt)/(2\pi )$%
\begin{equation}
emf_{ext}=2\pi RE_{ext}(R)=L\frac{di}{dt}=L\left( \frac{Ne}{2\pi }\frac{%
d^{2}\phi }{dt^{2}}\right)  \label{e22}
\end{equation}%
giving the self-inductance $L$ of the multi-charge ring system (where we can
ignore the negligible mass contribution) from Eq. (\ref{e20}) as%
\begin{equation}
L=\frac{2\pi RE_{e}(R)}{[Ne(d^{2}\phi /dt^{2})/(2\pi )]}=\frac{\left( 2\pi
\right) ^{2}R}{2c^{2}N}\sum\limits_{j=1}^{j=N-1}\left( \frac{\cos (2\pi j/N)%
}{|2\sin (\pi j/N)|}+\frac{\sin ^{2}(2\pi j/N)}{|2\sin (\pi j/N)|^{3}}\right)
\label{e23}
\end{equation}%
We see that the self-inductance of this multiparticle, circular-orbit
circuit is now independent of the mass $m$ or the charge $e$ of the current
carriers.\cite{carrierN} \ 

\subsubsection{Magnetic Energy of the Current Carriers}

The magnetic energy $U_{mag}=$ $\tint d^{3}r\mathbf{B}^{2}/(8\pi )$ stored
in the circular-orbit circuit is given by the cross-terms (but not the
self-terms) when the magnetic field is squared, and corresponds to the
velocity-dependent double sum in the Darwin Lagrangian Eq. (\ref{e4}). \
Thus we have%
\begin{eqnarray}
U_{mag} &=&\frac{1}{2}\tsum\limits_{i=1}^{N}\tsum\limits_{j\neq i}\frac{1}{%
8\pi }\tint d^{3}r\,2\mathbf{B}_{ei}(\mathbf{r})\cdot \mathbf{B}_{ej}(%
\mathbf{r})  \notag \\
&=&\frac{1}{2}\tsum\limits_{i=1}^{N}\tsum\limits_{j\neq i}\frac{e^{2}}{2c^{2}%
}\left( \frac{\mathbf{v}_{i}\cdot \mathbf{v}_{j}}{|\mathbf{r}_{i}-\mathbf{r}%
_{j}|}+\frac{\mathbf{v}_{i}\cdot (\mathbf{r}_{i}-\mathbf{r}_{j})\mathbf{v}%
_{j}\cdot (\mathbf{r}_{i}-\mathbf{r}_{j})}{|\mathbf{r}_{i}-\mathbf{r}%
_{j}|^{3}}\right)  \notag \\
&=&N\tsum\limits_{j=1}^{N-1}\frac{e^{2}}{2c^{2}}\left( \frac{v\widehat{y}%
\cdot \mathbf{v}_{j}}{\left\vert \widehat{x}R-\mathbf{r}_{j}\right\vert }+%
\frac{v\widehat{y}\cdot (-\mathbf{r}_{j})\,\mathbf{v}_{j}\cdot (\widehat{x}R)%
}{\left\vert \widehat{x}R-\mathbf{r}_{j}\right\vert ^{3}}\right)
\label{ee24}
\end{eqnarray}%
where in the last line of Eq. (\ref{ee24}) we have used $\mathbf{v}_{i}\cdot 
\mathbf{r}_{i}=0$ and have taken advantage of the symmetry to evaluate the
magnetic energy when the $N$th particle is located on the $x$-axis at $%
\mathbf{r}_{N}=\widehat{x}R$ and is moving with velocity $\mathbf{v}_{N}=%
\widehat{y}v=\widehat{y}Rd\phi /dt.$ The $j$th particle is located at $%
\mathbf{r}_{j}=R[\widehat{x}\cos (2\pi j/N)+\widehat{y}\sin (2\pi j/N)]$
with velocity $\mathbf{v}_{j}=R(d\phi /dt)[-\widehat{x}\sin (2\pi j/N)+%
\widehat{y}\cos (2\pi j/N)].$ \ Introducing these expressions along with the
distance given in Eq. (\ref{e17a}), the magnetic energy of Eq. (\ref{ee24})
is%
\begin{eqnarray}
U_{mag} &=&N\tsum\limits_{j=1}^{N-1}\frac{e^{2}}{2c^{2}}\left( R\frac{d\phi 
}{dt}\right) ^{2}\left( \frac{\cos (2\pi j/N)}{|2\sin (\pi j/N)|}+\frac{\sin
^{2}(2\pi j/N)}{|2\sin (\pi j/N)|^{3}}\right)  \notag \\
&=&\frac{1}{2}\left[ \frac{\left( 2\pi \right) ^{2}R}{2c^{2}N}%
\sum\limits_{j=1}^{j=N-1}\left( \frac{\cos (2\pi j/N)}{|2\sin (\pi j/N)|}+%
\frac{\sin ^{2}(2\pi j/N)}{|2\sin (\pi j/N)|^{3}}\right) \right] \left( 
\frac{eN}{2\pi }\frac{d\phi }{dt}\right) ^{2}  \label{e25}
\end{eqnarray}%
\qquad We recognize the current $i=eN(d\phi /dt)/(2\pi )$ and so can read
off the self-inductance of the circuit from $U_{mag}=(1/2)Li^{2}.$ \ The
expression for the self-inductance $L$ is the same as in Eq. (\ref{e23}). \
Now the power $P=eE_{ext}(R)v_{i}$ delivered to the $i$th charge by the
external electric field $\mathbf{E}_{ext}$ associated with the original emf
does not go into particle kinetic energy but rather is converted into
electromagnetic energy associated with the particles on the ring. \ This
electromagnetic energy is exactly the stored magnetic energy given by the
expression $(1/2)Li^{2}.$ \ 

For our example of an external emf acting on a circular charged-particle
circuit, the situation with many particles is totally transformed from the
situation with only one particle. \ As charged particles are added to the
circuit, the mechanical inertia increases linearly with the number of
particles $N$ while the inertia associated with the mutual electromagnetic
interactions increases quadratically with $N$. \ In the multiparticle case,
the mutual interaction between the particles overwhelms the single-particle
behavior so that the mass and charge of the individual charge carriers is no
longer of significance. \ The particles move so that the sum of the induced
electric fields at each particle cancels the external electric field.\cite%
{four}

\subsection{Discussion}

\subsubsection{Summary of the Calculations}

In the analysis above, we have discussed the self-inductance of a very
simple circuit from an unfamiliar point of view. \ We have evaluated the
electric fields of individual electric charges and shown how a system
involving a single charge is transformed over to a familiar electromagnetic
system when the number of charges is increased. \ Our simple example
involves charges under centripetal constraining forces giving a circular
orbit but allowing tangential acceleration along the circular orbit. \ In
the one particle example, the induced electric field depends upon both the
charge $e$ and the mass $m$ of the charge carrier with the induced field
proportional to $e^{2}/m.$ \ In the one-particle case, the energy
transferred to the particle goes into the kinetic energy of the ring
particle. \ However, when we deal with a multi-particle case, then the
electromagnetic forces between the charges are such as to transform the
behavior over to the familiar behavior of a conducting circuit where the
charge and mass of the current carriers are of no significance. \ When there
are a large number of charged particles, the acceleration of each charge is
determined by the requirement that the sum of the acceleration fields of all
the other charges should cancel the external electric field which produces
the external emf around the circuit. \ 

\subsubsection{Comments on "Hidden Mechanical Momentum"}

We suggest that the simple example discussed here is not only an interesting
approach to ideas of self-inductance, but also is a reminder that it is
unlikely that effects of particle mass will be of importance in
multiparticle electromagnetic systems. \ Some forty years ago, the claim was
made that relativistic effects of \textit{mechanical} particle momentum were
needed to account for the conservation of linear momentum in the
poorly-understood interaction of a magnet and a charged particle.\cite{CVV}
\ This perplexing magnet-charge interaction is the basis for controversies
associated with the Aharonov-Bohm effect,\cite{AB} the Aharonov-Casher
effect,\cite{AC} and Mansuripur's paradox.\cite{Mansuripur} \ The associated
idea of \textquotedblleft hidden mechanical momentum\textquotedblright\ has
become entrenched in the research and textbook literature of
electromagnetism.\cite{Griffiths}\cite{Jackson}\cite{letter} \ However, the
only valid calculation of \textquotedblleft hidden mechanical
momentum\textquotedblright\ involves particles which have negligible mutual
interactions; these calculations are equivalent to a one-particle magnet.%
\cite{calc} \ All of the other cases of \textquotedblleft hidden mechanical
momentum\textquotedblright\ are simply empty claims without valid
calculations. \ In the present transparent analysis, the one-particle
example where particle inertia dominates is completely different from the
mutually-interacting multiparticle limit where electromagnetic inertia
dominates. The one-particle example contributes no understanding and has no
relevance to the multiparticle case; it simply misleads physicists. \ We
suggest that the idea of \textquotedblleft hidden mechanical
momentum\textquotedblright\ is an error based on invalid extrapolations from
a one-particle example. \ We believe that \textquotedblleft hidden
mechanical momentum\textquotedblright\ is extraordinarily unlikely to have
any relevance as a valid physical idea within multiparticle electromagnetism.


\begin{thebibliography}{99}
\bibitem{unaware} I am not aware of any such treatment carried out in the
literature.
\bibitem{carrierN} It seems interesting that for this circular charge
arrangement, the typical multiparticle behavior requires at least four
charges. \ The summation in Eq. (\ref{e23}) is steadily increasing with
increasing particle number $N$. \ However, the summation starts out negative 
$(-0.5)$ for $N=2,$ and is still negatiave $(-0.289)$ for $N=3.$ \ Only for $%
N=4$, does the summation become positive $\ (+0.207).$
\end{thebibliography}
\end{document}